# Device Platform for Electrically Reconfigurable Dielectric Metasurfaces

Prasad P Iyer,[1] Mihir Pendharkar,[1] Chris J. Palmstrøm[1,2] and Jon A. Schuller[1*]
[1]Electrical and Computer Engineering Department, University of California Santa Barbara, Santa Barbara, CA-93106
[2]Material Science and Engineering Department, University of California Santa Barbara, Santa Barbara, CA-93106
*Corresponding author: jonschuller@ece.ucsb.edu

**ABSTRACT** Achieving an electrically tunable phased array optical antenna surface has been a principal challenge in the field of metasurfaces. In this letter, we demonstrate a device platform for achieving reconfigurable control over the resonant wavelength of a subwavelength optical antenna through free-carrier injection. We engineer and grow, using molecular beam epitaxy, a heterostructure of $In_{1-x}Al_xAs/InAs/Al_yGa_{1-y}Sb$ layers designed to achieve large amplitude and phase modulation of light by maximizing the refractive index change in regions of resonant field enhancement The p-i-n layers are grown on a heavily doped n-InAs layer which forms a reflecting substrate to confine the Mie resonances within the nanowires of the index tunable layers. We outline the fabrication process developed to form such tunable metasurface elements using a four-step projection lithography process and a self-aligned vertical dry etch. We experimentally demonstrate the operation of an electrically reconfigurable optical antenna element where the resonant wavelength blue shifts by 200nm only during carrier-injection. We extrapolate the experimentally measured InAs refractive index shifts to show we can achieve nearly π phase shift in a metasurface array. This solid-state device platform enables us to contact each resonant element independently to form a truly reconfigurable Fourier optical element with the promise of arbitrary control of the electromagnetic wavefront at the subwavelength scale.

## INTRODUCTION

Metasurfaces have great ability to transform input plane waves into desired output beamforms upon passing through an ultrathin resonant optical antenna surface [1]. Researchers have exploited subwavelength control over the local phase [2], amplitude [3] and polarization [4] of light to achieve high efficiency static metasurface lenses [5], deflectors [6], Bessel [7] and vortex beam [4] transformers etc. These static components highlight the versatility of metasurfaces and point to the great application potential afforded by reconfigurable metasurfaces. Attempts to achieve dynamic metasurface control exploit a variety of methods and materials including: i) phase change materials [8], ii) 2D material [9] or 2D electron gas [10,11] based plasmonic metasurfaces, iii) Insulator to metal phase transitions [12], iv) thermo-optic effects [13], v) thermal [14] or optical [15] free-carrier effects vi) liquid crystal [16] based tuning vii) MEMS [17] based tuning. Each approach has varying application-specific advantages and disadvantages but share in common the underlying ability to actuate large modulation of a resonant wavelength [18]. Theoretically proposed [19] free-carrier modulated Mie-resonator metasurfaces are particularly promising for high-speed, electrically controlled operation, but lack experimental demonstration. In this letter, we demonstrate large resonance wavelength tuning via free-carrier accumulation in InAs metaresonator diodes.

The most common approach for electrically-reconfigurable metasurfaces relies on depletion-based devices that modulate high carrier density two-dimensional electron gasses [9–11]. Because of inherent trade-offs between depletion width and carrier density, such devices necessarily operate near highly absorbing epsilon near zero (ENZ) wavelengths. As an alternative, we previously proposed InSb/InAlSb p-i-n heterojunction Mie resonators operating in forward bias (i.e. carrier accumulation) as a basic building block for reconfigurable metasurfaces [19]. These resonators must sit on highly reflecting back electrodes. To avoid difficulties in integrating MBE grown layers on metallic substrates through flip-chip bonding [20], we develop here a monolithically integrated solution using heavily doped InAs layer ($5\times10^{19} cm^{-3}$ of 1.25μm thickness) as the back-electrode reflector. The doped InAs layer forms a highly reflective substrate (>90% reflectivity for λ>8μm) on which we grow an $In_{1-x}Al_xAs/InAs/Al_yGa_{1-y}Sb$ p-i-n heterojunction.

## DEVICE HETEROSTRUCTURE

The device heterostructure, grown by Molecular Beam Epitaxy (MBE), is designed to maximize the free-carrier based refractive index change in an intrinsic InAs region during forward bias. A heterojunction p++GaSb/i-InAs/n++InAs p-i-n stack (Fig 1A) is grown atop a heavily doped n++ InAs back reflector. The n++InAs/p++GaSb heterojunction forms a tunneling ohmic contact. Hence, we form a common contact to all of the devices using the heavily doped InAs layer. Starting with a Te doped GaSb substrate, a buffer layer of GaSb was grown, followed by the n++ InAs layer (grown at 2μm per hour), acting as the back reflector. This n++ InAs layer is assumed to be relaxed on the GaSb substrate and the heterostructure above has been designed with the in-plane lattice parameter of InAs as the host layer. A p-type i-AlSb/p-AlGaSb superlattice minimizes electron-hole recombination via tunneling from the i-InAs conduction band to the p++GaSb valence band during forward bias. Due to the large conduction band offset ($\Delta E_c$ =1eV) with InAs, the AlGaSb/AlSb superlattice also acts as an electron blocking layer (EBL) during forward bias (fig 1B). The $Al_{0.5}Ga_{0.5}Sb$ layers are p-type doped up to $2\times10^{18} cm^{-3}$ to ensure minimum resistance for hole injection while alternate layers of i-AlSb layers are introduced to maximize the electron resistance. The average valence band maximum in the p-type superlattice is designed to match the i-InAs valence band maximum. Ultimately, the i-AlSb/p-AlGaSb superlattice prevents direct band-to-band tunneling while minimizing the forward bias required for hole injection. The 1.25μm unintentionally doped (UID) InAs layer grown above the p-type EBL acts as an index tunable layer where the resonant electromagnetic fields are primarily confined. Hole-blocking layers (HBL) are grown above the i-InAs using an alternating n-$InAl_xAs_{1-x}$/i-InAs superlattice with the %Al (x) concentration increasing as we grow the structure (Fig 1A). This ensures i) these layers have a minimal lattice mismatch with InAs and preferably do not relax on the InAs, ii) $\Delta E_v$ keeps increasing to provide larger resistance to hole extraction and iii) the n-type dopants of InAlAs ensure minimum resistance to electron injection during forward bias. The whole structure is capped with a layer of n++ InAs to form an ohmic contact. *In-situ* Reflection High Energy Electron Diffraction (RHEED) during the MBE growth of InAlAs/InAs strained layers, and on termination of growth, showed no immediate signs of lattice relaxation and maintained a 2x4 reconstruction (Fig 3A). RHEED intensity oscillations were also observed throughout the growth of the strained layer.

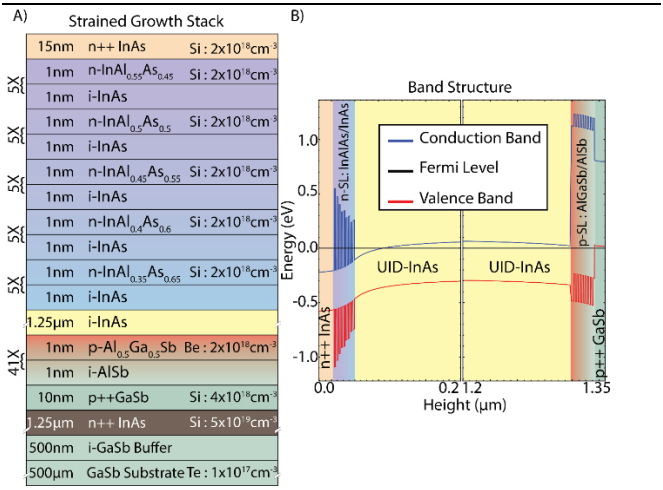

Figure 1: A) Device heterostructure, as grown on a GaSb substrate. Each layer is color-coded w.r.t the band structure plot in panel B, with the layer thickness written on the left and doping concentration on the right side within each block. B) The band structure showing the conduction band (blue), valence band (red) and fermi-level (black) at equilibrium state. The structure highlights the electron-blocking (right) and hole-blocking (left) layers along with the highly doped ohmic contact layers on either end of the device.

The same metal stack can be used to form contact to both the p- and n-type sides of the diodes due to the band offsets between InAs and GaSb. We have combined both numerical Poisson solvers [21] and full wave finite difference time domain (FDTD) simulations [22] to optimize this InAs based device stack with an EBL (i-AlSb/p-Al$_y$Ga$_{1-y}$Sb) grown directly on a highly doped back reflector and an HBL (n-In$_{1-x}$Al$_x$As/i-InAs) that is strained to the 1.25μm i-InAs. This unique structure enables us to tune the refractive index of the 1.25μm i-InAs layer due to electron and hole accumulation, in steady state conditions, during forward bias operation.

## FABRICATION PROCESS

We employ a novel fabrication process that includes four projection lithography steps on an auto-stepper lithography tool. The device design is optimized within the fabrication limits and alignment tolerances of the available stepper (optical projection) lithography system.

**A) Metal Antenna Deposition and Lift-off**: 120nm metal stacks (Mo/Ni/Au/Pt/Ti: 10/5/55/10/40nm) are deposited with e-beam evaporation, patterned into split-gap electrodes using a positive photoresist (SPR-955) and negative polarity mask, then lifted off the top side of the bonded epitaxial layer (Fig 2A) using a heated N-Methyl-2-pyrrolidone (NMP) solution at 80°C for 4 hours. The metal stack is designed to a) minimize contact resistance to InAs (Mo/Au), b) restrict metal diffusion into the epi-layer during current injection or thermal processing (Mo) c) provide good adhesion between Mo and Au (Ni) & d) to act as a hard-mask during the subsequent mesa etch step (Pt/Ti). The Ti layer forms a sacrificial protective layer for the metals as well as a hard mask for etching, and is finally removed after step D using a buffered HF (BHF) wet etch dip. Ideally, the metal top electrode minimally perturbs the optical modal confined primarily within the i-InAs. Hence, the presence of sharp edges due to lift-off wings would dramatically affect the device performance. The photoresist (PR: SPR 955CM) verticality defines this critical feature, and is accentuated using a contrast enhancing absorbing layer above the PR and an adhesion promoter (Hexa-Methyl-Di-Silazane, HMDS) below the PR. The epi-layers are dipped in NH$_4$OH solution for 10s before metal deposition to minimize the native oxide formation on the III-V layers.

**B) Oxide Hard Mask**: A positive PR is patterned to protect the gap in antenna/contact layer lifted of in the previous step (Fig 2B) and a 350nm SiO$_2$ hard mask is then conformally deposited using a plasma enhanced chemical vapor deposition (PECVD) process at 250°C. Lithography alignment errors less than 50nm was achieved through iterative calibration steps. The stepper's specified 250nm auto-alignment error was circumvented by calibrating the misalignment of the mask/lens system using a fine-Vernier test structure. The oxide hard-mask is dry etched using an inductively coupled plasma (ICP) with CHF$_3$ and CF$_4$ gases at 50W radio-frequency (RF) power. The gas flow ratios were optimized from a standard recipe to operate at lower power while maintaining a vertical etch. The lower power minimizes the sputter damage to the metals and semiconducting epi-layer. After the oxide etch, the residual PR is removed using an acetone, isopropyl alcohol and water clean.

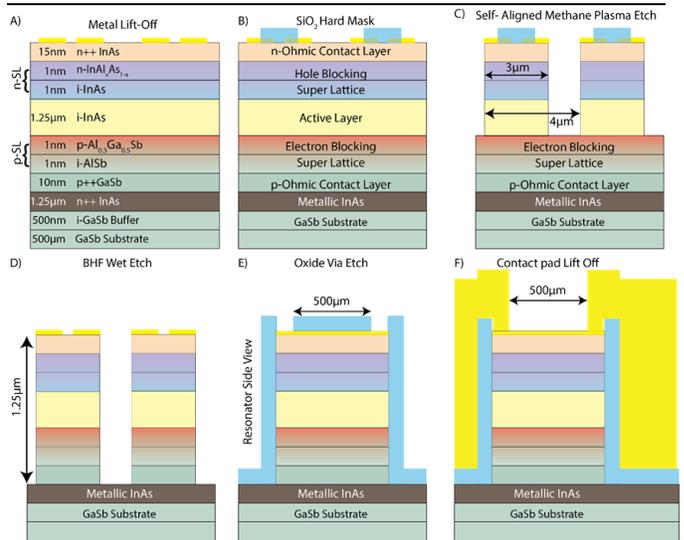

**Figure 2:** Panels A-F demonstrate the fabrication process flow developed to form the tunable heterojunction metasurface

**C & D) Mesa Etch**: Reactive Ion etching (RIE) in a CH$_4$/Ar/H$_2$ plasma chamber is used to achieve vertical and relatively smooth sidewalls for InAs alloys (Fig 2C). The SiO$_2$ and Ti layers are used as the hard-mask, providing >10:1 etch selectivity for etching the III-V alloys. For this etch, the Al-containing EBL epi-layers provided a natural etch stop. Due to long etch times (∼ 30 mins), we developed a Bosch-like process to minimize build-up of polymer etch residues. The residual hard masks (Ti and SiO2) and Al-containing etch stop layers are removed using a BHF wet etch (Fig 2D), exposing the n++InAs layer. Typical meta-resonator defined at this stage are approximately 1 micron tall, 3 microns wide, and 1000 microns long.

**E) Oxide Mask for Contact Vias**: After the BHF wet etch, 350nm PECVD SiO$_2$ is deposited atop a patterned positive PR to define metal contact pads at the ends of the ∼ 1000 micron long meta-resonators (Fig 2E). The low power oxide etch in step B (Fig 2B) is repeated. This conformal oxide coating prevents the shorting of the top and bottom contact.

**F) Contact Pad Lift-Off**: A bi-layer of positive PR (1.2μm) and negative SF11 (PMGI) resist is used to define the lateral contact pad dimensions. Contact pads comprising 1.7μm Au and 300nm Pt are e-beam evaporated and then lifted off, followed by dry etching the SiO$_2$ deposited in the previous step (Fig 2F). SEM images of the contacts made to the resonators are shown in the Fig 4A for single optical antenna resonators. This multi-step fabrication process defines individually contacted one-dimensional tunable resonators.

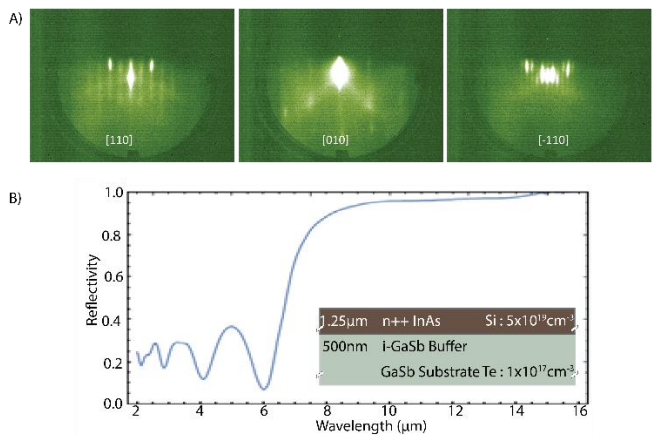

**Figure 3: A)** 2x4 reconstruction on the surface of strained n-InAs (001) observed at the end of growth of p-i-n diode stack. **B)** The reflectivity curve plotted as a function of the wavelength from the doped InAs substrate (shown in the inset) demonstrating the back-reflecting mirror it forms at long wavelengths.

## RESULTS AND DISCUSSIONS

This work focusses on individual heterojunction meta-resonator diodes to demonstrate a blue shift in the resonant wavelength during carrier-injection. Reflection measurements (normalized to the substrate reflectance) of single meta-resonator diodes are measured in an FTIR microscope at room temperature for different applied voltages. The reflectance curve of the doped InAs exposed after the fabrication process enables us to prove that the substrate doping forms a metallic substrate (Fig 3B) at mid-infrared wavelengths with reflectance values > 95% for λ>10μm and >90% for λ>8μm. The plasma frequency edge of the reflection curve at 6μm shows that the refractive index of InAs layer is consistent with Drude model after accounting for the non-parabolic electron effective mass in InAs [23].

While the p-i-n heterostructure is designed to be strained to the underlying n++ InAs layer, the 0.6% lattice mismatch between the n++ InAs and the GaSb substrate leads to a high density of dislocations that propagate through the device heterostructure. Uniformly spread, electrically conducting defects and dislocations currently limit device yield and subsequent fabrication of large area meta-surface arrays. Further work on growing a fully lattice matched structure, including a lattice matched n++ back reflector layer, is required for scaling up from individual meta-resonators. We have previously proposed a theoretical device heterostructure for alleviating issues arising out of high density of dislocations, by moving to a lattice matched InSb materials platform [19]. A typical I-V curve for a single meta-resonator diode is plotted in Fig 4C. The device exhibits a turn on voltage of 0.2V, matching the simulated band structure (valence band) offsets (Fig 1B). The normalized reflection measurements reveal a fundamental TE resonance near 13.5μm. The resonance wavelength shifts to lower wavelengths by 200nm during forward bias, with negligible shift during reverse bias (Fig 4D). The blue-shift, which is in the opposite direction from thermo-optic effects due to Joule heating, is produced due to free-carrier injection and accumulation during forward bias. According to electromagnetic simulations, this 200nm blue shift corresponds to a change of ~0.2 in the real part of the i-InAs refractive index and an associated free-carrier accumulation of $4.5 \times 10^{17} cm^{-3}$ [23]. From these single resonator results, we estimate the expected phase shift of metasurface arrays using FDTD simulations. The simulated reflection amplitude and phase are shown in Figures 5B and 5C respectively. The reflection amplitude exhibits a resonance dip due to absorption within the n++ InAs back reflector. Compared to single resonators, the array linewidth is approximately 2.3 times smaller due to reduced radiative broadening [24]. As such, our observed 200nm shift produces up to a π phase shift (Fig 4c) with 0.5V of applied bias.

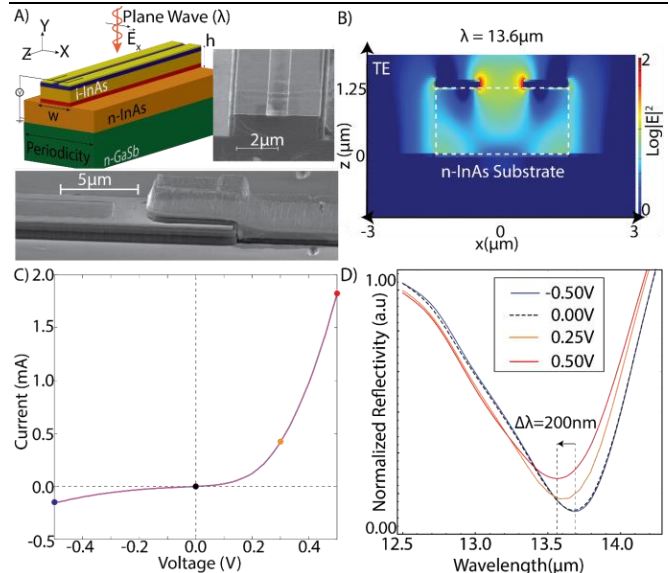

**Figure 4: A)** Sketch of a device and the SEM image of the device cross-section at the end of the process cycle. Side view of a single optical antenna device showing the split-dipole antenna contacted from the side. **B)** The resonant electric field intensity within the resonator is plotted at 13.6μm. **C)** Measured device characteristics of a single optical antenna showing the diode operation with a turn on voltage near 0.25V matching with the expected band-alignments. The color dots corresponds to the curves in the adjacent panel. **D)** The normalized reflection curves measured from a single optical antenna at different applied voltages (black dashed at 0V, blue at -0.5V, yellow at 0.25V and red at 0.5V) corresponding the color dots on the panel A. The dip in reflection shows the voltage dispersion of the resonant wavelength of the antenna which blue shifts only during forward bias operation. The curves are vertically shifted to elucidate the wavelength shift with applied bias.

In summary, we demonstrate a heterojunction meta-resonator diode whose resonance wavelength can be dynamically tuned through free-carrier injection. The demonstrated results show promise for constructing reconfigurable metasurface elements, although further development is needed to increase tunability and yield. Reducing the density of dislocations is critical for achieving uniform large-area metasurface arrays. It was also observed that better surface passivation post-fabrication enables larger free-carrier accumulation due to reduced side-wall leakage and carrier recombination. Lastly, a higher quality back reflector electrode is expected to further reduce on-resonance absorption losses. This work paves the way for realization of solid-state electrically reconfigurable metasurfaces.

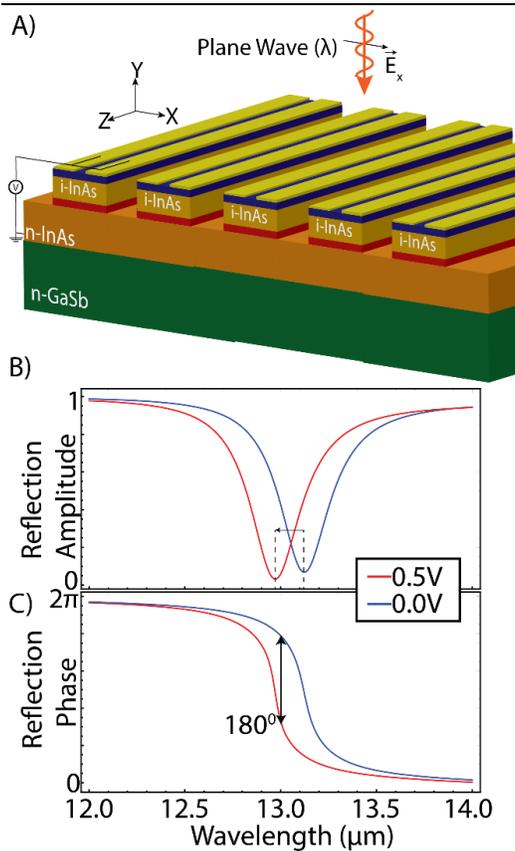

**Figure 5**: A) Sketch showing an array (periodicity = 5μm, w=4μm, h=1.25μm and g=1μm) of InAs resonators B and C) Reflection amplitude (B) and Phase (C) of the simulated array showing the blue-shift of the resonant wavelength at 0V (blue line) and 0.5V (red line) which causes 180° phase shift at 13.3μm.

**Funding:** This work was supported by the Air Force Office of Scientific Research (FA9550-16-1-0393). We also acknowledge support from the Centre for Scientific Computing from the CNSI and NSF CNS-0960316. The authors would like to acknowledge the support of the UCSB MRL Shared Experimental Facilities which are supported by the MRSEC Program of the NSF under Award No. DMR 1121053; a member of the NSF-funded Materials Research Facilities Network. M.P. and C.J.P. would like to acknowledge support from the Vannevar Bush Faculty Fellowship program sponsored by the Basic Research Office of the Assistant Secretary of Defense for Research and Engineering and funded by the Office of Naval Research through grant N00014-15-1-2845. A part of this work was performed in the UCSB Nanofabrication Facility which is a part of the NSF funded National Nanotechnology Infrastructure Network.